\documentclass[prl,aps,onecolumn,superscriptaddress, showpacs]{revtex4-1}

\usepackage{graphicx}
\usepackage{dcolumn}
\usepackage{bm}
\usepackage{ulem}
\usepackage{times}

\usepackage[utf8]{inputenc}
\usepackage[english]{babel}
\usepackage{keyval}
\graphicspath{{figs/}{figs:}{\figs}}
\setkeys{Gin}{width=0.90\columnwidth}

\usepackage[usenames,dvipsnames]{color}
\usepackage[]{ulem}

\newcommand{\comment}[1]{\textcolor{red}{#1}}
\renewcommand{\comment}[1]{\relax}

\newcommand{\todelete}[1]{\textcolor{green}{\sout{#1}}}
\renewcommand{\todelete}[1]{\relax}

\begin{document}

\title{Anomalous charge and negative-charge-transfer insulating state in cuprate chain-compound KCuO$_2$}

\author{D. Choudhury}
\email{debraj@phy.iitkgp.ernet.in}
\affiliation{Department of Physics, University of Arkansas, Fayetteville, AR 72701, USA}
\affiliation{Department of Physics, Indian Institute of Technology, Kharagpur 721302, INDIA}

\author{P. Rivero}
\affiliation{Department of Physics, University of Arkansas, Fayetteville, AR 72701, USA}

\author{D. Meyers}
\affiliation{Department of Physics, University of Arkansas, Fayetteville, AR 72701, USA}

\author{X. Liu}
\affiliation{Department of Physics, University of Arkansas, Fayetteville, AR 72701, USA}

\author{Y. Cao}
\affiliation{Department of Physics, University of Arkansas, Fayetteville, AR 72701, USA}

\author{S. Middey}
\affiliation{Department of Physics, University of Arkansas, Fayetteville, AR 72701, USA}

\author{M. J. Whitaker}
\affiliation{Department of Chemistry and Chemical Biology, Rutgers University, Piscataway, NJ 08854-8087, USA}

\author{S. Barraza-Lopez}
\affiliation{Department of Physics, University of Arkansas, Fayetteville, AR 72701, USA}

\author{J. W. Freeland}
\affiliation{Advanced Photon Source, Argonne National Laboratory, Argonne, Illinois 60439, USA}

\author{M. Greenblatt}
\affiliation{Department of Chemistry and Chemical Biology, Rutgers University, Piscataway, NJ 08854-8087, USA}

\author{J. Chakhalian}
\affiliation{Department of Physics, University of Arkansas, Fayetteville, AR 72701, USA}

\begin{abstract}

Using a combination of X-ray absorption spectroscopy experiments with first principle calculations, we demonstrate that insulating KCuO$_2$ contains Cu in an unusually-high formal-3+ valence state, the ligand-to-metal (O to Cu) charge transfer energy is intriguingly negative (${\Delta}$ $\sim$ $-$1.5 eV) and has a dominant ($\sim$60$\%$) ligand-hole character in the ground state akin to the high Tc cuprate Zhang-Rice state. Unlike most other formal Cu$^{3+}$ compounds, the Cu 2$\it{p}$ XAS spectra of KCuO$_2$ exhibits pronounced 3${d}$$^8$ (Cu$^{3+}$) multiplet structures, which accounts for $\sim$40$\%$ of its ground state wave-function. ${Ab~initio}$ calculations elucidate the origin of the band-gap in KCuO$_2$ as arising primarily from strong intra-cluster Cu 3$\it{d}$ - O 2$\it{p}$ hybridizations ($\it{t}_{\rm{pd}}$); the value of the band-gap decreases with reduced value of $\it{t}_{\rm{pd}}$. Further, unlike conventional negative charge-transfer insulators, the band-gap in KCuO$_2$ persists even for vanishing values of Coulomb repulsion $\it{U}$, underscoring the importance of single-particle band-structure effects connected to the one-dimensional nature of the compound.

\end{abstract}

\date{\today}
\maketitle

\noindent The electronic properties of strongly correlated transition metal (TM) oxides $-$which consist of partially filled TM ${d}$-orbitals hybridized with the ligand (oxygen) ${p}$-orbitals$-$ are effectively categorized under the well known Zaanen-Sawatzky-Allen (ZSA) phase diagram \cite{ZSawatzkyAPRL1985,AEBocquetPRB1992,AEBocquet1996}, a guiding principle for materials scientists that takes into consideration the on-site ${d}$-${d}$ Coulomb interaction energy at the TM site (${U}$) and the ligand-to-TM charge transfer energy (${\Delta}$). There is an intriguing region of the ZSA phase diagram of compounds with negative values of ${\Delta}$ that has been less explored \cite{JSSCDD1990,AFujimori1991,PramanaDD1992,PRBDD1993,AFujimori1994,YTokura1998,DIKhomskii2011}. In TM oxides the value of ${\Delta}$ decreases by increasing the valence (oxidation) state of the TM ion, and for unusual high valence states ${\Delta}$ can even become negative \cite{AFujimori1994}. Such high-valence compounds are very unstable, and only a few pristine negative ${\Delta}$ compounds exist (see Table \ref{3tabstddielectrics}). For such highly covalent compounds, it is energetically favorable to transfer an electron from the ligand to the metal ion, as the energy cost ${\Delta}$ for this process is negative, giving rise to a large ${ligand}$-${hole}$ character and usually metallic nature of the ground state. However, there exists a very select number of compounds, which are insulating while having negative or extremely small values of ${\Delta}$, driven by a combination of strong metal-ligand hybridization either with electronic correlations, which are known as the correlated covalent insulators \cite{JSSCDD1990,PramanaDD1992,PIASDD1994}, as in La$_2$CuO$_4$ \cite{PRBDD1989}, Sr$_2$CuO$_3$ \cite{EPLAFujimori1997} or with single-particle band-structure effects, as in NaCuO$_2$ \cite{AFujimori1991,AFujimori1994,PIASDD1994,PRBDD1993,DJSingh1994,DIKhomskii1997}.

In this work, using X-ray absorption spectroscopy (XAS) experiments, model XAS and density functional theory (DFT) calculations, we have investigated the electronic structure of KCuO$_2$ \cite{EKaiserThA1995}, and have elucidated the nature of its experimentally observed insulating state. Our results show that KCuO$_2$ hosts Cu in a formal 3+ valence state, has a negative ${\Delta}$ and a dominant ligand-hole character on its ground state. We find a charge band gap ($\sim$ 1.24 eV) with preponderance of O 2$\it{p}$ states at the valence band and conduction band edges, which originates from strong intra-cluster Cu 3$\it{d}$ - O 2$\it{p}$ hybridization in this negative ${\Delta}$ compound and competes with point-charge Coulomb contributions to the crystal-field energies of the Cu $\it{t}_{2\rm{g}}$ orbitals. The chain topology driven band-gap persists for vanishing ${U}$, which  is distinct from the conventional  picture of correlated covalent insulators \cite{JSSCDD1990,PramanaDD1992,PIASDD1994}, and also decreases with decreasing values of $\it{t}_{\rm{pd}}$. The inclusion of strong correlations is, however, necessary to account  for the experimental value of  the gap.  Our work thus establishes that KCuO$_2$, similar to NaCuO$_2$, is a negative ${\Delta}$ insulator where the insulating behaviour arises from both single-particle band structure effects  from the unique one-dimensional CuO$_2$ chain geometry and strong electron-electron correlations.
\begin{table}
  \centering
  \caption{Coulomb repulsion ${U}$ and charge-transfer ${\Delta}$ energies (in units of eV) for some transition metal oxides with unusually high formal-valence states for the $\it{B}$ site (Fe, Co, Ni, Cu) cation.}
  \label{3tabstddielectrics}
  \vspace*{0.6cm}
\begin{tabular}{|c|c|c|c|c|c|}
  \hline
  \hline
   Compound& Formal-valence & ${U}$ & ${\Delta}$ & Tranport & Ref.  \\
  \hline
  SrFeO$_{3}$ &(4+)& 7.8 & 0.0 & Metal & \cite{AEBocquetPRB1992}\\
  BaFeO$_{3}$ &(4+)& 7.1 & $-$0.9 & Insulator & \cite{PRBHWadati2015}\\
  SrCoO$_{3}$ &(4+)& 7.0 & $-$5.0 & Metal & \cite{GSMAbbate1995} \\
  LaNiO$_3$ &(3+)& 7.0 & 1.0 & Metal & \cite{MValletRegi2002}\\
  LaCuO$_{3}$ &(3+)& 7.0 & $-$1.0 & Metal & \cite{AFujimori1998}\\
  NaCuO$_{2}$ &(3+) & 8.0 & $-$2.5 & Insulator  & \cite{PRBDD1993}\\
  KCuO$_{2}$ &(3+) & 8.0 & $-$1.5 & Insulator  & this work\\
  \hline
  \hline
\end{tabular}
\end{table}

\noindent{}${Methods~(experimental)}.-$ Polycrystalline KCuO$_2$ in a single-phase orthorhombic $\it{CmCm}$ space group \cite{RHoppe1969} was synthesized by mixing KO$_2$ and CuO powders in 1:1 ratio in Ar-filled glovebox followed by sintering under a dry O$_2$ atmosphere for 2.5 days at 450$^o$C \cite{DRLines1989}. XA measurements at Cu ${L}$$_{3,2}$- and O ${K}$-edges were performed on the 4-ID-C beam line of the Advanced Photon Source (APS) at Argonne National Laboratory, USA. The sample powder was mounted on the holder using carbon-tape under nitrogen gas atmosphere to ensure minimum exposure to air, and XAS measurements in total-electron-yield (TEY), total-fluorescence-yield (TFY),  and in the inverse-partial-fluorescence-yield (IPFY) modes were performed at room temperature without any additional surface preparation.  The probing depth in case of the TEY ($\sim$5 nm) is much smaller than that of TFY or IPFY  ($\sim$100 nm)\cite{TVVenkatesan2013}, and, thus, while TEY studies the under-coordinated surface electronic-structure of a solid, the TFY and IPFY are well-suited to investigate the bulk electronic-structure. For IPFY measurement, the non-resonant O $\it{K}$-edge was monitored, and, thus, IPFY is further free from any self-absorption effects unlike TFY \cite{DGHawthorn2011}.

\noindent{}${Methods~(theory)}.-$ We have performed three sets of complementary calculations. To act as reference XAS spectra, calculations of the Cu ${L}$$_{3,2}$ XA spectrum on a orthorhombic ${Cmcm}$ space-group lattice of KCuO$_2$ \cite{RHoppe1969} (e.g., Figs. 1(a-b)) were performed using the Finite Difference Method Near-Edge Structure (FDMNES) code \cite{YJoly2009}. The FDMNES calculations were performed using the full-multiple-scattering theory with a cluster radius of 6 \AA ~ around the absorbing Cu atom and an on-site Coulomb energy (${U}$) of 8 eV.

In order to determine the relative TM-O covalencies, cluster calculations for simulating the Cu ${L}$$_{2,3}$ XA spectrum of a single CuO$_2$ planar-cluster with a ${D}$$_{4h}$ symmetry \cite{CommentClusterXAS} were performed using the Charge Transfer Multiplet program for X-ray Absorption Spectroscopy (CTM4XAS) \cite{StavitskiFMFdeGroot2010}. The charge transfer energy ${\Delta}$ between Cu 3${d}$ and O 2${p}$ orbitals is defined as ${E}$(${d}^{n+1}\underline{{L}}$)$-$${E}$(${d}^{n}$), where ${E}$(${d}^{n}$) is the multiplet-averaged energy for ${n}$-electron occupancy on Cu 3${d}$ levels and ${E}$(${d}^{n+1}\underline{{L}}$) denotes the multiplet-averaged energy obtained after transferring one electron from an O 2${p}$ level to the Cu 3${d}$ level having ${n}$=8 electrons, corresponding to the formal (3+) valence state of Cu. For the CTM4XAS calculations, the basis size was restricted up to one electron charge-transfer from O 2${p}$ to Cu 3${d}$.

To determine the density of states (DOS) of KCuO$_2$ and NaCuO$_2$, the rotationally invariant LDA+U scheme of Dudarev ${et ~ al.}$ \cite{Dudarev} was employed in DFT electronic structure calculations. Calculations were carried out with the ${Vienna ~ Ab ~ initio ~ Simulation ~ Package}$ (VASP) \cite{Kresse} using projector-augmented wave pseudopotentials \cite{Blochl,Kresse2}. The first Brillouin zone was sampled using a 12$\times$12$\times$6 Monkhorst-Pack set of k-points and a 400 eV energy cutoff.


\begin{figure}[htp]
\includegraphics[width=0.48\textwidth]{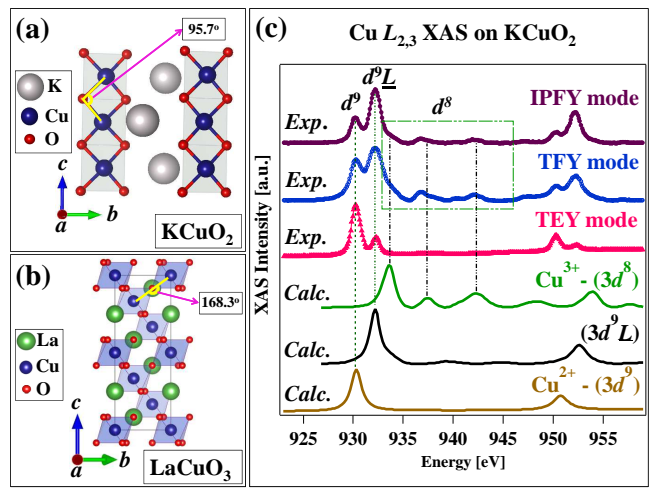}
\caption{\label{} (Color online) Schematic crystal structures showing (a) edge-sharing of CuO$_4$ units in KCuO$_2$ and (b) corner-sharing of CuO$_6$ clusters for LaCuO$_3$. (c) Cu ${L}$$_{2,3}$ X-ray absorption (XA) spectra of KCuO$_2$ collected in the inverse-partial-fluorescence-yield (IPFY), the total-fluorescence-yield (TFY) and the total-electron-yield (TEY) modes. Calculated XA spectra (solid lines) of KCuO$_2$ with the FDMNES code for 3${d}^9$, 3${d}^8$ and 3${d}^9$$\underline{{L}}$ configurations are also shown with the experimental spectra.}
\end{figure}

\noindent{}{\it Results and discussion}$.-$ Two distinct groups of experimentally-observed XAS peaks, one around 930 eV (${L}$$_3$ region) and another group around 950 eV (${L}$$_2$ region) can be clearly identified for the Cu ${L}$$_{3,2}$-edge of KCuO$_2$ (c.f., Fig.~1(c)). While the spectral features for the ${L}$$_3$ and ${L}$$_2$ regions are nearly identical, they are separated by about 20 eV due to the 3/2 $\times$ Cu 2${p}$ core-spin-orbit coupling. Looking around the Cu ${L}$$_3$ region closely, we observe two intense peaks at 930.3 eV and 932.3 eV, which correspond to the Cu ${d}^9$ and the Cu ${d}^9$$\underline{{L}}$ initial states, respectively \cite{PRB1988-DD,EISolomon2006,JChakhalian2013}.

The ${d}^9$ (Cu$^{2+}$) peak intensity increases significantly in the TEY mode as compared to the TFY and IPFY modes, indicating an abundance of Cu$^{2+}$ valence states on the surface (see Methods section). This Cu$^{2+}$ presence is believed to arise due to the presence of surface impurity phases rich in Cu$^{2+}$. Note that similar peaks of ${d}$$^9$ (Cu$^{2+}$) have been observed for other formally Cu$^{3+}$ compounds in the XA spectrum (e.g., NaCuO$_2$ \cite{PRB1988-DD,AFujimori1991,AFujimori1994}, CaCu$_3$Co$_4$O$_{12}$ \cite{JChakhalian2013} and Cs$_2$KCuF$_6$ \cite{FMFdeGroot1998}). Cu$^{2+}$ impurity phases on the surfaces of these metastable compounds arises due to the loss of superficial anionic atoms during XAS experiments in ultra-high vacuum which effectively reduces the valence of surrounding Cu ions \cite{AFujimori1994}. Further, we observed that KCuO$_2$ on exposure to air decomposes into CuO within five to ten minutes. Thus, given these constraints, it is impossible for us to avoid the Cu$^{2+}$ related impurity peak in the XAS experiments. Within the bulk, KCuO$_2$ is not expected to suffer from such anionic losses and, accordingly, much lower intensity Cu$^{2+}$ peaks in the bulk-sensitive TFY and IPFY XAS spectra are observed in Fig. 1(c). Some percentage of the TFY and IPFY signals are also contributed from the surface and near-surface region of the sample, which is dominant due to powder nature of KCuO$_2$ sample as compared to scraped bulk-polycrystalline pellet of NaCuO$_2$ \cite{PRB1988-DD}, that still provides significant contributions of the ${d}$$^9$ peak. Since TFY, unlike IPFY, suffers from self-absorption effects \cite{DGHawthorn2011}, it causes the differences in their relative spectral weights.

\indent Focussing henceforth on the IPFY spectrum, as it is both bulk-sensitive and free from self-absorption effects, the main peak given by the ${d}^9$$\underline{{L}}$ state arises due to the charge transfer of an electron from the surrounding O atoms into formally Cu 3${d}^8$ (Cu$^{3+}$) state \cite{PRB1988-DD,EISolomon2006,JChakhalian2013}. Furthermore, distinct multiplet-structures --that are considered to provide clear evidence for the presence of an ionic Cu$^{3+}$ (${d}$$^8$) state \cite{PRB1988-DD,EISolomon2006,JChakhalian2013}-- are observed around 940 eV. The presence of significant ${d}^9$$\underline{{L}}$ and ${d}^8$ intensities suggests that a coherent superposition of both states constitutes the ground state of formal Cu$^{3+}$ ions in KCuO$_2$, similar to that of NaCuO$_2$ \cite{PRB1988-DD}. It is important to note that on a Cu 2${p}$-3${d}$ XAS process it is difficult to detect contributions from the ${d}^{10}$$\underline{{L}}$$^2$ level to the ground-state. However, such contributions are usually small, as determined by X-ray photo-electron spectroscopy on related systems \cite{AFujimori1994}.

\indent To further establish the origin of the various features in the experimental XAS spectra, we simulated the Cu ${L}$$_{3,2}$ XAS spectra of KCuO$_2$ that corresponds to the ${d}$$^8$, ${d}^9$ and ${d}^9\underline{{L}}$ initial state configurations using the FDMNES code. As shown by the vertical guide lines in Fig. 1(c), the calculated XAS spectra corresponds to the ${d}$$^9$ and ${d}$$^9\underline{{L}}$ features in the experimental spectra, and the observed ionic ${d}^8$- experimental features can be broadly understood with the calculated spectrum for the ${d}$$^8$ ionic Cu$^{3+}$ state.


\indent We now compare the ${L}$$_3$ energy region for KCuO$_2$ with other systems that host unusual valence states of Cu, such as  the optimally-doped YBa$_2$Cu$_3$O$_{7-\delta}$ (YBCO) \cite{JChakhalian2013}, LaCuO$_3$ \cite{JChakhalian2013}, and NaCuO$_2$ \cite{PRB1988-DD} in Fig. 2(a), after subtraction of the surface Cu$^{2+}$ impurity peak \cite{DCPeakSubtraction2014,SpectraNaCuO2DD}. It is interesting to note that the Zhang-Rice spin-singlet-state, ${d}^9$$\underline{{L}}$ \cite{GASawatzky1988,ZhangRice1988}, which arises due to external hole-doping in YBCO by intricate chemical routes \cite{PRB1988-DD}, naturally becomes the dominant state in formally Cu$^{3+}$ compounds. This hole-doping mechanism is akin to a self-doping effect \cite{DIKhomskii2014}.
Judging from the intensity ratios shown in Fig. 2, the ${d}^9$$\underline{{L}}$ charge-transfer state appears dominant over the ionic ${d}^8$ state for KCuO$_2$, NaCuO$_2$ and LaCuO$_3$, thus suggesting that the associated charge transfer energies ${\Delta}$ for all of these compounds are unusually $\it{negative}$. We note that negative values of ${\Delta}$ have been already proposed for insulating NaCuO$_2$ \cite{AFujimori1991,AFujimori1994,PRBDD1993} and metallic LaCuO$_3$ \cite{AFujimori1998}.

\begin{figure}[htp]
\includegraphics[width=0.48\textwidth]{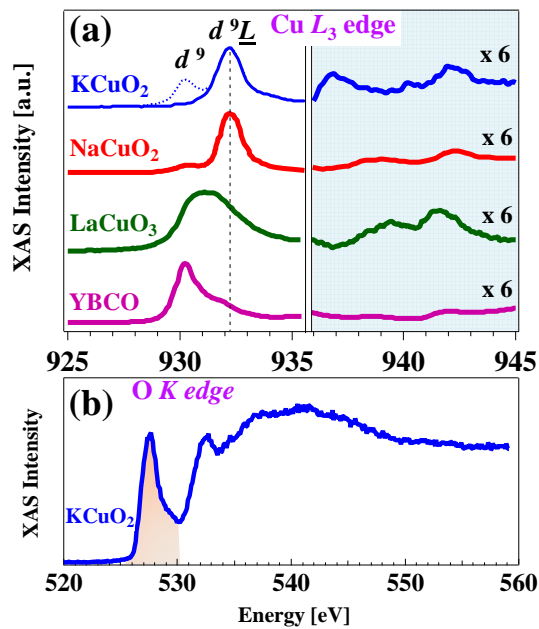}
\caption{\label{} (Color online) (a) Cu ${L}$$_{3}$ X-ray absorption (XA) spectra of KCuO$_2$, NaCuO$_2$ \cite{PRB1988-DD}, LaCuO$_3$ and YBa$_2$Cu$_3$O$_{7-\delta}$ (YBCO). The main Cu ${L}$$_{3}$ peak in KCuO$_2$ and NaCuO$_2$, and the shoulder in YBCO around 932 eV correspond to the ${d}^9$ $\underline{{L}}$ Zhang-Rice singlet state. The ${d}^8$ multiplet structures six-fold-increased for easier observation are also shown. (b) The O ${K}$-edge XA spectrum of KCuO$_2$ consists of a pronounced pre-peak around 527.6 eV (shaded area) suggesting a large ligand-hole character of its ground state.}
\end{figure}

\indent A closer analysis of the XA shapes on Fig. 2(a) points to spectral differences within several formal Cu$^{3+}$ compounds. Let us focus on the differences in the XA spectral features related to the ${d}$$^9\underline{{L}}$ state first: The ${d}$$^9\underline{{L}}$ peak for LaCuO$_3$ is broad and can be well described using two peaks, one centered at 930.8 eV and another at 932.2 eV. This splitting occurs from the delocalization of the ligand-hole, due to inter-cluster hybridization effects that are aided by the corner-sharing geometry of the CuO$_6$ clusters with Cu-O-Cu bond angle of 168.3$^\circ$ in LaCuO$_3$ \cite{AFujimori1998} (c.f., Fig.~1(b)). For KCuO$_2$ and NaCuO$_2$, on the other hand, such inter-cluster hybridization effects are negligible due to the near-orthogonal Cu-O-Cu bond-angle (95.7$^\circ$) between neighboring CuO$_4$ clusters (c.f., Fig. 1(a)) and a single ${d}$$^9\underline{{L}}$ peak is observed.

\indent The ${d}^8$ multiplet region of formally Cu$^{3+}$ compounds shown by the shaded area in Fig. 2(a) is discussed next. Covalency and ${\Delta}$ are not independent, since the relative intensities between the ${d}^8$ multiplets to the ${d}^9$$\underline{{L}}$ peak usually increase with decreasing covalency, and their energy separation increases with larger negative values of ${\Delta}$ \cite{FMFdeGroot1998}. KCuO$_2$ has stronger multiplet intensities than iso-structural NaCuO$_2$, which suggests a larger contribution of the ionic ${d}^8$ state to its ground state.  Further, the average energy difference between the ${d}^8$ multiplets and the ${d}^9$$\underline{{L}}$ peak is 5.9 eV and 8.2 eV for KCuO$_2$ and NaCuO$_2$, respectively, thus showing a smaller negative ${\Delta}$ for KCuO$_2$.

\indent For the calculated Cu ${L}$$_{3,2}$ XA spectra on a single CuO$_2$ cluster with planar ${D}$$_{4h}$ symmetry, we optimized the parameter values to match the calculated energy separations between the average ${d}$$^8$ multiplets and the ${d}$$^9\underline{{L}}$ main peak with energy differences obtained from experiment (Fig.~2(a)). The estimated ${\Delta}$, thus obtained, turned out to be $-$1.5 eV and $-$2.5 eV for KCuO$_2$ and NaCuO$_2$ respectively. Furthermore, both the resultant ground states have dominant ${d}$$^9\underline{{L}}$ characters, 39$\%$${d}$$^8$ + 61$\%$${d}$$^9\underline{{L}}$ (36$\%$${d}$ $^8$ + 64$\%$${d}$$^9\underline{{L}}$) for KCuO$_2$ (NaCuO$_2$), with a higher ionic character for the ground state of KCuO$_2$, as suggested earlier.


\begin{figure}{\includegraphics[width=0.48\textwidth]{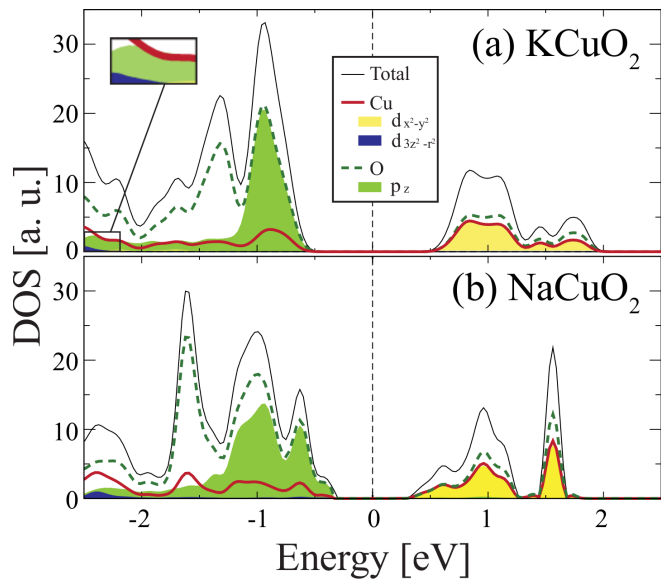}}
\caption{(Color online) Density of states for (a) KCuO$_2$ and (b) NaCuO$_2$; the total contributions from a given atomic species are indicated by trendlines and the orbital projections are shown by colored area plots. A ${U}$ of 8 eV was used for these calculations. KCuO$_2$ is found to exhibit a larger bandgap than NaCuO$_2$.}
\end{figure}

\indent The O ${K}$-edge XA spectrum $-$which probes the ligand hole-states$-$ exhibits a pronounced pre-peak for KCuO$_2$ at 527.6 eV, as seen in Fig.~2(b). The intensity of the O ${K}$-edge pre-peak correlates directly with the amount of ligand-hole character in the ground state \cite{FMFdeGroot1998}, thus the strong pre-peak in KCuO$_2$ further establishes a large ${d}$$^9$$\underline{{L}}$ character of its ground state.

Fig.~3 shows the density of states (DOS) projected onto orbital contributions for KCuO$_2$ and NaCuO$_2$, which were found to have insulating gaps of 1.24 eV and 0.62 eV respectively for the ${U}$ value of 8 eV. The band-gaps in KCuO$_2$ and NaCuO$_2$, however, exist even for $U =$ 0 eV, in agreement with previous observations on NaCuO$_2$ \cite{PRBDD1993,PIASDD1994,DJSingh1994}, highlighting the role of single-particle band-structure effects due to  the chain topology in giving rise to the insulating state in KCuO${_2}$. The inclusion of correlations, however, is essential in increasing the band-gap value as compared to $U =$ 0 eV and bringing it to the agreement with the experimental value \cite{AFujimori1994}. Furthermore, the projected DOS shows a strong O character in both valence and conduction band edges \cite{O2pOrbitalLDA}. The Cu $\it{t}$$_{\rm{2g}}$ levels occur between the lower-lying 3$\it{d}$$_{3z^2-r^2}$ and higher-lying 3$\it{d}$$_{x^2-y^2}$ levels, as usually observed for one-dimensional CuO$_2$ chains due to point charge (Coulomb) contribution \cite{CrystalField}. However, the $\it{t}$$_{\rm{2g}}$ levels is intriguingly seen to have Cu ($\it{d}_{xz}$) and Cu($\it{d}$$_{yz}$) character immediately below $\it{E}$$_{\rm{F}}$ and Cu $\it{d}$$_{xy}$ character only at further lower energies, which is different from a point charge (Coulomb) contribution to crystal-field splitting. Similar effect has been observed in Cs$_2$Au$_2$Cl$_6$, and arises from a dominant ${pd}$ covalency contribution in case of negative $\Delta$ compounds \cite{DIKhomskii2011}; the inversion of the $\it{t}_{\rm{2g}}$ orbitals thus further confirms the negative $\Delta$ in KCuO$_2$.

We also performed a Bader analysis \cite{RefBader1990} to understand the charge density distribution over electronic orbitals. The total occupation of Cu 3${d}$-shell in both systems is 8.8, which represents a mixture of ${d}$$^8$ and ${d}$$^9$ states, in qualitative agreement with cluster calculations and establishing the superposition of both contributions to the ground state of formally Cu$^{3+}$ ions in KCuO$_2$ and NaCuO$_2$, as discussed earlier.

\noindent{}${Conclusions.-}$  We have described the presence of the anomalous charge state of Cu in KCuO$_2$ from experiment and theory. We established the negative charge transfer energy of the KCuO$_2$ ground state and its dominant ligand-hole character, which arise due to large intra-cluster hybridization effects and remain localized due to weak inter-cluster hybridizations. Localized cuprate like Zhang-Rice singlet state thus occur at every unit cell, which consequently gives rise to the experimentally observed insulating and diamagnetic character of KCuO$_2$ \cite{EKaiserThA1995,DIKhomskii2014}. Moreover, KCuO$_2$ exhibits strong ${d}$$^8$ related multiplet structures, resulting from the large ionic Cu$^{3+}$ character of its ground-state. KCuO$_2$ is shown to belong to the unusual class of covalency driven negative charge transfer with the correlated gap that  is adiabatically connected  to the single-particle gap arising the chain geometry of the compound.

We deeply thank  D. D. Sarma and  D. I. Khomskii for  insightful suggestions and comments. We  thank NSF-XSEDE (Grant TG-PHY090002; TACC's {$Stampede$) and HPC at Arkansas for computational support. This research at the University of Arkansas is funded in part by the Gordon and Betty Moore Foundation's EPiQS Initiative through Grant GBMF4534 and by the DOD-ARO under Grant No. 0402-17291. P.R. and S.B.L. acknowledge funding from the Arkansas Biosciences Institute.

\end{document}